# Influence of the potential barrier switching frequency on the effectiveness of energy harvesting

Jerzy MARGIELEWICZ[1], Damian GĄSKA[1], Grzegorz LITAK[2], Tomasz HANISZEWSKI[1], Piotr WOLSZCZAK[2], Carlo TRIGONA[3]

[1]*Silesian University of Technology, Faculty of Transport and Aviation Engineering, Krasińskiego 8, Katowice 40-019, Poland*

[2]*Lublin University of Technology, Faculty of Mechanical Engineering, Nadbystrzycka 36, Lublin 20-618, Poland*

[3]*Dipartimento di Ingegneria Elettrica Elettronica ed Informatica, University of Catania, 95125 Catania, Italy*

Corresponding author

Damian GĄSKA

Silesian University of Technology, Faculty of Transport and Aviation Engineering, Krasińskiego 8, Katowice 40-019, Poland, damian.gaska@polsl.pl

# Influence of the potential barrier switching frequency on the effectiveness of energy harvesting

**Abstract:** Dynamical systems for kinetic energy harvesting have recently been designed that use nonlinear mechanical resonators and corresponding transducers to create electrical output. Nonlinearities are important for providing a broadband input frequency efficiency, which is necessary to ensure that the output voltage is at a satisfactory level in the case of a variable ambient vibration source. In this paper, we analyze the dynamics of such a system with the additional possibility of a switching potential barrier in the nonlinear resonator. In order to improve the effectiveness of the energy harvester, we numerically test the possibility of using abrupt changes in the system parameters. We analyze the voltage output for various frequencies of both the harmonic excitations and the potential barrier switching. In a range that includes low and high switching frequencies of potentials, the periodic solutions dominate in the bifurcation diagrams. With regard to the large switching frequency, the effective values of the voltage at the piezoelectric electrodes are significantly reduced. The highest effective values of the voltage that are induced at the piezoelectric electrodes are observed when the chaotic solutions appear due to their specific ability to pass the potential barrier for relatively small excitation. The results also show that mechanical damping with high-enough magnitudes minimize the influence of the transient states caused by the switching potentials.

## 1. Introduction

Small size vibration energy harvesters that transform ambient vibration energy to electrical power were proposed, in terms of linear devices, a significant amount of time ago [1,2]. Considering the density of energy, the piezoelectric transducers were optimized for their size between one to tens of centimeters [3]. However, the variation in the ambient conditions cause such a linear device to operate outside of resonance, which leads to a considerable reduction in the power output [4]. The most popular way to increase this output is to apply nonlinear effects to the resonator. In such a system, the frequency broadband effects were observed [5–7]. One of the frequency amplitude diagrams, which they are related to, is modified in the region of the resonance and there is also the appearance of some new classes of resonances in the form of rational fractions or multipliers of the linearized system natural frequency [8–11]. The additional effects are based on the multiple solutions, which depends on the initial conditions that are very common in the nonlinear systems and the appearance of non-periodic (chaotic) solutions [12,13]. The nonlinear potential could be modeled by springs [14–16] and magnets [17,18]. For the second option, the positions and orientations of the magnets are very important factors and lead to single or multiple potential wells.

Typical kinetic energy harvesters for obtaining energy from vibrations are usually constructed from a housing in which a cantilever beam is mounted, and a piezoelectric transducer is glued on it [19–21]. Systems with several potential wells were designed to better use the possibilities of obtaining energy from non-linear systems. And so, systems with bistable (Bistable Energy Harvesters BEH), tristable (TEH) and even quadstable (QEH) and pentastable (PEH) characteristics are used [22–25]. Many types of such multistable systems have been studied both theoretically and in laboratory experiments. The bistable system was tested e.g. in the work of Derakhshani et al. [26] where parameters including vibrational motion, output voltage, and frequency response were analyzed both theoretically and experimentally under different excitation conditions. Similar tests were performed for different designs [27–29]. However, the greatest attention of researchers so far has focused on TEH systems, which have been widely investigated [30–32]. Litak et al. presented the results, focused on identifying multiple solutions, of numerical simulations of a non-linear tristable system for harvesting energy based on permanent magnets [33].

Undoubtedly, the amount of recovered energy and the effectiveness of energy harvesters still pose a challenge [18].

Zhou et al. [34] modified the typical design solution of TEH by changing the inclination angle of two external magnets, so that the ability to change the angle of rotation enabled the improvement of energy efficiency. Such a design change made it possible to obtain various shapes of the potential function. Constructions with moving magnets have also been featured in other works. Xinxin et al. [35] proposed an energy harvester built of a piezoelectric cantilever beam with a tip magnet and a movable magnet connected by a spring to the housing, which broadened operating bandwidth and improved harvesting power. Zhou et al. [36] had the same goal when he introduced a bi-stable energy harvester based on two flexible beams with a variable potential energy function. Changing the position of magnets to change the potential function has also been proposed in other works . In all cases, the variable function was intended to improve energy efficiency [18,37,38]. On the other hand, at the vicinity of the critical excitation amplitude, competing multiple solution can lead to chaotic response accompanied by passing through the potential barrier [13]. The appearance of chaotic and transient chaotic solutions could be more important for additional parametric excitation leading to variable potential barriers. Such variability represent additional way of system excitations and simultaneously provide more solutions. In this context identification of specific solutions could be more important [39]. In this work we consider an important case of simultaneous inertial and parametric excitations. All these studies motivated us to propose a design solution based on non-smooth dynamics – abrupt change of state of the potential function as for switches, impacts etc.

In the present paper, we investigate the effects of potential switches during the harvesting work. We also analyze the voltage output for various frequencies of both the harmonic excitations and the potential barrier switching for BEH, TEH and system with non-smooth dynamics. Diagrams of the RMS values of the voltage induced at the piezoelectric electrodes are compared to the corresponding bifurcation diagrams. To identify particular solutions, we used the phase portraits, Poincaré maps, and Fourier spectra. In Section 3 we present the results of the influence of parameters like: the excitation amplitude, the sequence of potential barriers switching and its frequency on the dynamics of the system. Finally, the solutions are classified by the corresponding average voltage output.

## 2. Mathematical model formulation

The subject of the model tests presented in this paper is an energy harvesting system with a cyclically switched potential barrier (Fig. 1). Here, the tested structure is composed of a flexible cantilever beam *I*, which is fixed in a nondeformable body *III* and attached via bolts *IV* to the vibrating housing of the object. Under the influence of the mechanical vibrations described by the harmonic function $q_1 = Asin(\omega_W t)$, beam *I* is precipitated from the equilibrium position. As a result, a voltage is induced on the piezoelectric electrodes *II*. The tested energy harvesting system has the ability to assume dynamic states, which are mapped by two potential barriers. At a given moment in time, the operating point of the system is mapped with the potential of one barrier. The subsystem consists of a nondeformable *V* cylinder with permanent magnets mounted onto its external surfaces, which are responsible for the cyclical change of the nonlinear characteristics. To minimize the impact of the magnetic fields, the inside of the cylinder is filled with *VI* material, whose task is to separate the magnetic fields that are generated by permanent magnets. The cyclical change in the angular orientation of the *V* cylinder causes the potential barrier to be mapped with either a two-well (Fig. 1b) or three-well (Fig. 1c) characteristics.

The influence of inertia and resistance to the movement of the *V* cylinder is negligible, meaning that the potential barrier can be switched over a very short period of time. From a theoretical point-of-view, the simplest mathematical description of the switching potential barriers is to map it through a rectangular time sequence. We realize that this is a far-reaching idealization of the dynamics of the switching circuit. However, adopting this simplification provides new qualitative and quantitative information. It is important to note that the technical solution of the switching system is not discussed in this paper. Idealized characteristics that reflect the shape of the potential, and its influence, were mapped for any moment in time by means of cyclograms. The color of the cyclogram (Fig.1) is directly correlated with the color of the potential barrier. The times $t_i$, highlighted in the cyclograms, define the change in angular orientation of the *V* cylinder.

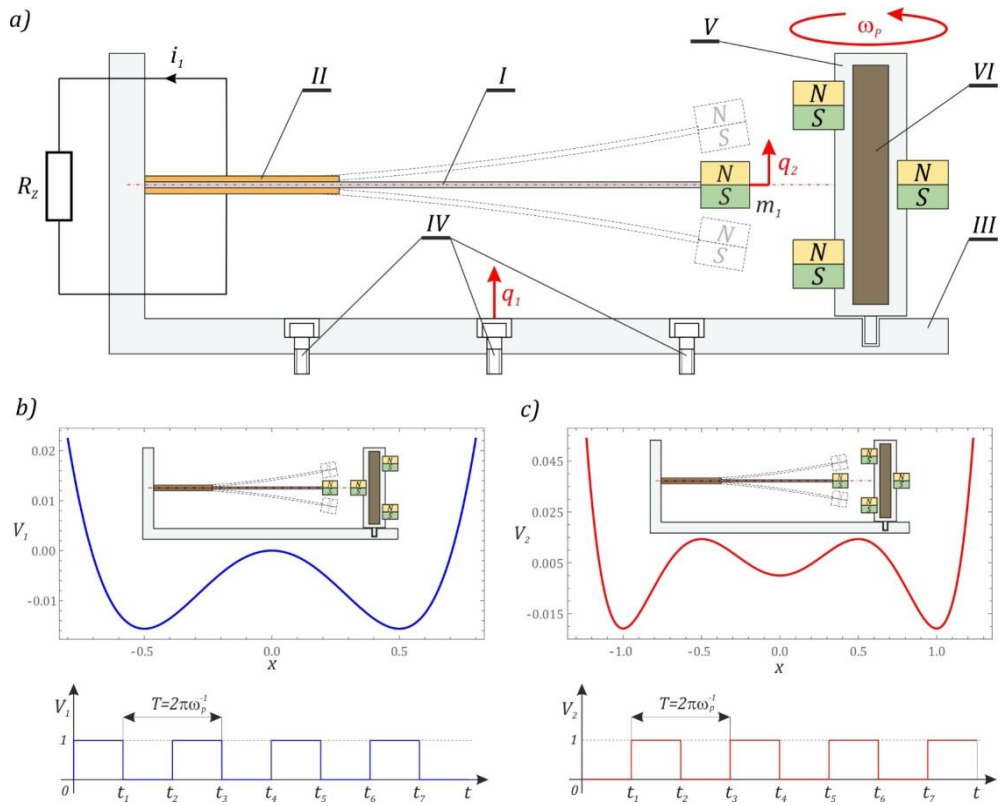

**Fig. 1.** a) Schematic diagram of the energy harvesting system with cyclically switched potential barrier. b), c) certain configurations of the magnets and the corresponding potential barriers.

The permanent magnets mounted onto the cylinder were selected in such a way that the minimums of the two-well barrier (Fig. 1b) coincide with the local maximums of the barrier, which were initiated by the two permanent magnets (Fig. 1c). Here, the mechanical characteristics reflecting the stiffness of an elastic cantilever beam is included in the modeling functions of the cyclically changing potentials. Based on the formulated phenomenological model, the differential equations of motion were derived, which constitute the formal basis for conducting the quantitative and qualitative numerical experiments. These are written as:

$$\begin{cases} \dfrac{d^2y}{dt^2} + \dfrac{b_B}{m}\dfrac{dy}{dt} + (c_{12}y^3 - c_{11}y)\dfrac{F_{S1}}{m} + (c_{23}y^5 - c_{22}y^3 + c_{21}y)\dfrac{F_{S2}}{m} + \dfrac{k_P}{m}u = -\dfrac{d^2q_1}{dt^2}, \\ \dfrac{du}{dt} + \dfrac{u}{C_P R_Z} - \dfrac{k_P}{C_P}\dfrac{dy}{dt} = 0. \end{cases} \quad (1)$$

Table 1 describes each coefficient of Eq. 1, furthermore, the variable $y$ represents the difference in the displacements of the movable magnet mounted at the free end of the flexible cantilever beam $q_2$; the kinematic excitation is caused by the vibrations of the object body $q_1$ from which the energy is harvested. The values $F_{S1}$ and $F_{S2}$ model the potential barrier of the switching circuit, and their time characteristics are described with time sequences that pulsate from zero. From a mathematical point-of-view, the data control functions are algebraic equations, and their graphical images are plotted in graphs (Fig. 1), so that:

$$F_{S1} = \tfrac{1}{2} + \tfrac{1}{2}sgn[sin(\omega_P t)], \quad F_{S2} = \tfrac{1}{2} + \tfrac{1}{2}sgn[sin(\omega_P t + \pi)]. \quad (2)$$

From a theoretical point of view, the characteristics that define the potential barrier can be identified through experimental research. Such an approach is efficient and effective, thanks to which reliable measurement data is obtained, which is necessary to carry out numerical experiments [32,40]. An alternative approach is theoretical considerations based on the laws and principles of mechanics and electrical engineering. The effects of such analyzes significantly depend on the assumed simplifications and the adopted idealization of the research object [36]. Much better identification results are achieved using the finite element method (FEM)[41,42]. At this point, it is worth mentioning that the results obtained by the finite element method allow graphically depicting magnetic field lines that are invisible to the naked eye during laboratory tests. Analysis of the magnetic field lines makes it possible to estimate the scale of interaction between permanent magnets. Figure 2 show exemplary graphical images of the distribution of magnetic field lines between the magnet loading the free end of the flexible cantilever beam and the magnets fixed in the cyclically rotating V-frame. The presented exemplary results of numerical experiments illustrate the distribution of magnetic field lines for various magnet configuration.

a)
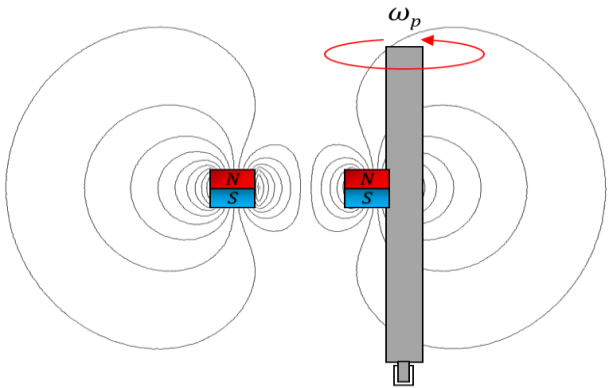

b)
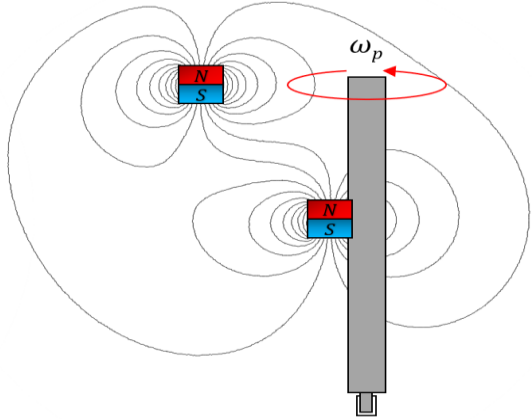

c)
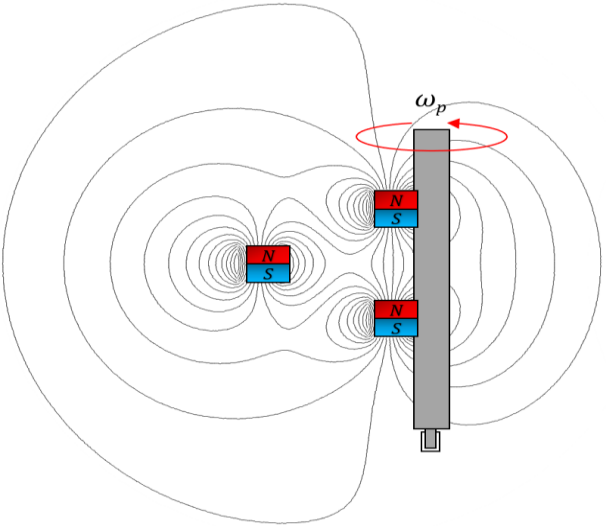

d)
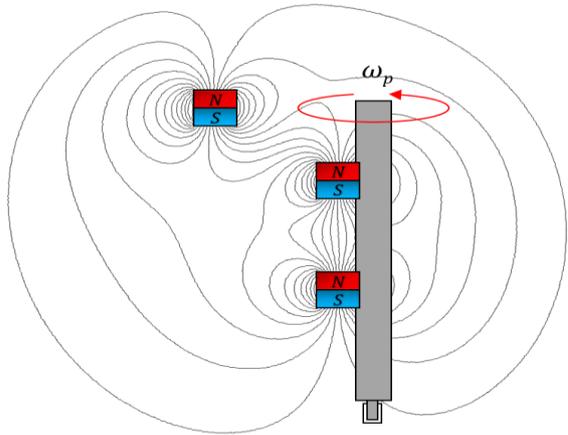

**Fig. 2.** Examples of magnetic flux lines (magnets interaction) for extreme magnet positions – sample data obtained by FEM, a), b) – bistable potential well, c), d) – tristable potential well. Simulation prepared in Finite Element Method Magnetics software

When conducting the numerical experiments, the control functions can be replaced with the *If* conditional. Nevertheless, it takes longer for the computer simulations to run because the conditional statement must be checked during each iteration. In the potential switching method that is to be proposed, this will not be necessary because we rely on continuous trigonometric functions. These functions take the values 0, when the potential is absent, and 1 for cases when the potential barrier is actively affecting the energy harvesting system. The parameter $\omega_P$, on the other hand, is responsible for the switching frequency of the potential barriers. The electric circuit factors represent, respectively, the constant $k_P$ and piezoelectric capacity $C_P$. The $R_Z$ parameter corresponds to the equivalent internal resistance of the piezoelectric and the external electrical circuit:

$$\begin{cases} \ddot{x} + \delta\dot{x} + \mu(\gamma x^5 - \beta x^3 + x)\left(\frac{1}{2} + \frac{1}{2}sgn[sin(\omega_S \tau + \pi)]\right), \\ +(\alpha x^3 - x)\left(\frac{1}{2} + \frac{1}{2}sgn[sin(\omega_S \tau)]\right) + \theta u = \omega^2 p sin(\omega\tau) \\ \dot{u} + \sigma - \vartheta\dot{x} = 0, \end{cases} \quad (3)$$

where

$$\mu = \frac{c_{21}}{c_{11}}, \quad \delta = \frac{b_B}{m\omega_0}, \quad \alpha = \frac{a_0^2 c_{12}}{c_{11}}, \quad \beta = \frac{a_0^2 c_{22}}{c_{21}}, \quad \gamma = \frac{a_0^4 c_{23}}{c_{21}}, \quad \theta = \frac{k_P}{ma_0\omega_0^2},$$

$$\vartheta = \frac{k_P a_0}{C_P}, \quad \sigma = \frac{1}{\omega_0 C_P R_Z}, \quad \omega_0^2 = \frac{c_{11}}{m}, \quad \omega = \frac{\omega_W}{\omega_0}, \quad \omega_S = \frac{\omega_P}{\omega_0},$$

$$x = \frac{y}{a_0}, \quad p = \frac{A}{a_0}, \quad \tau = \omega_0 t.$$

Using such a formulated mathematical model for the energy harvesting system with switched potential barriers, it was possible to perform the model tests.

## 3. Model test results and discussion

Numerical experiments, which map the dynamics of the energy harvesting system with switched potentials, were carried out with reference to the numerical data summarized in Table 1.

Table 1. Geometric and physical parameters of the model.

| Name | Symbol | Value | |
|---|---|---|---|
| Inertial element (mass) loading the beam | $m$ | 0.03 kg | |
| Energy losses in a mechanical system | $\delta$ | 0.138 Nsm$^{-1}$ | |
| | $c_{i1}$ | 20 Nm$^{-1}$ | 15 Nm$^{-1}$ |
| Parameters defining the potential barriers | $c_{i2}$ | 512×10$^3$ Nm$^{-3}$ | 480×10$^3$ Nm$^{-3}$ |
| | $c_{i3}$ | | 2.46×10$^9$ Nm$^{-5}$ |
| Scaling factor | $a_0$ | 0.0125 m | |
| Equivalent resistance of the electrical circuit | $R_Z$ | 1.1×10$^6$ Ω | |
| Equivalent capacity of the electric circuit | $C_P$ | 72 nF | |
| Electromechanical constant of piezoelectric converter | $k_P$ | 3.98·10$^{-5}$ N/V | |

For the first stage of the numerical experiments, the effectiveness of the energy harvesting of the system with a switched barrier was compared to solutions that are based on two and three wells. The formal basis for a comparison of the energy generation effectiveness involves diagrams of the effective values of the voltage that are induced at the piezoelectric electrodes in the steady state [5,43]. Based on preliminary computer simulations, it was established that the steady state of the energy harvesting system occurs after 450 periods of the source of the mechanical vibrations. On the other hand, the effective value of the voltage recorded on the piezoelectric electrodes was calculated in relation to the time sequences with a duration of 150 periods of the mechanical vibrations that affect the energy harvesting systems. On considering the direct comparison of the results obtained, the effective value diagrams were plotted that assume the same external load conditions and computer simulation settings. The presented results of model tests were obtained for zero initial conditions. For each identified diagram, the mean values were calculated in the individual bands for the variation of the dimensionless excitation frequency that affect the energy harvesting systems (Fig. 3).

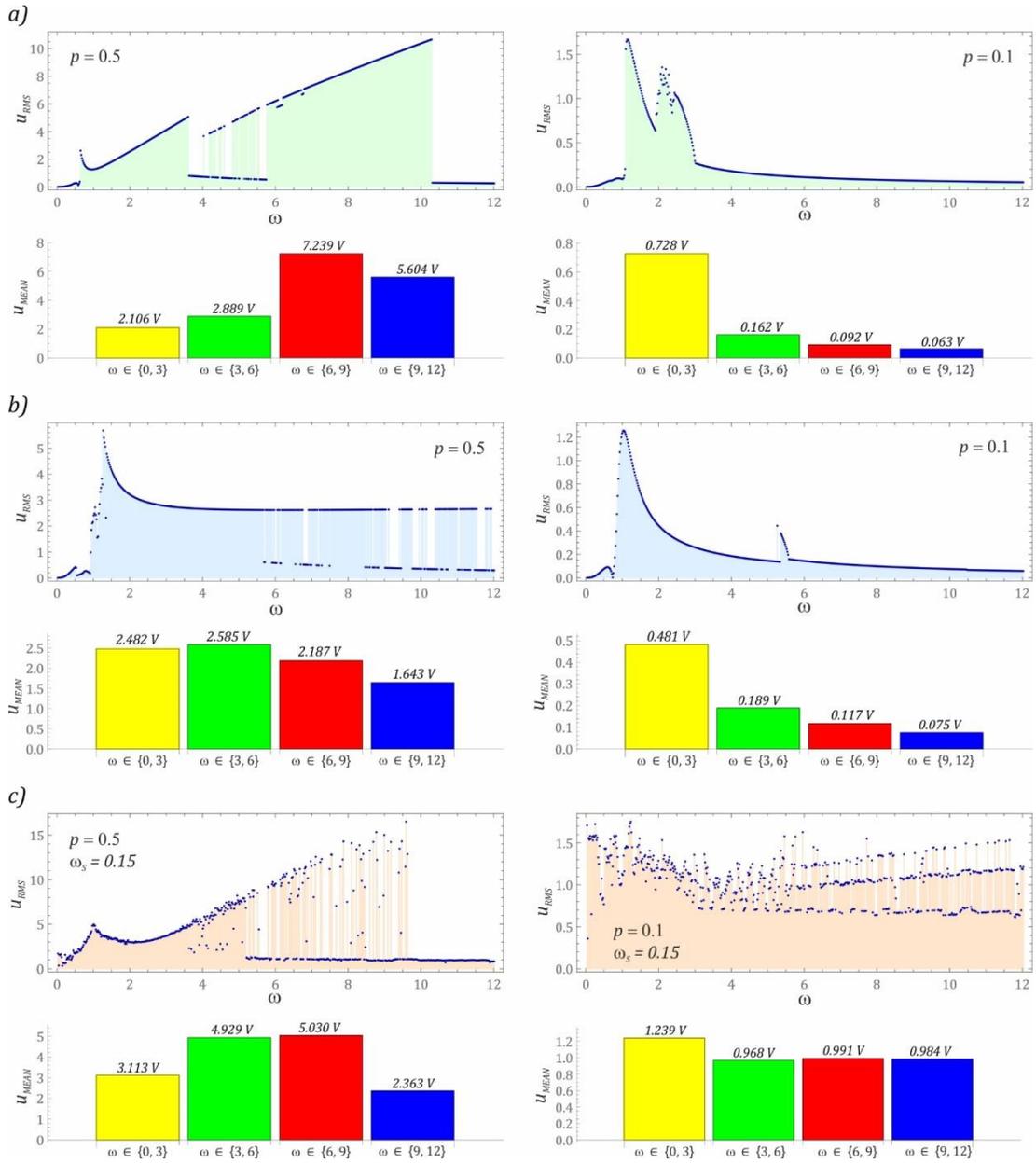

**Fig. 3.** Diagrams for the RMS values of the voltage induced at the piezoelectric electrodes: system with a) two wells, b) three wells, and c) a switchable potential barrier. Relevant parameters are indicated in each figure.

It is clear that the average value is a subjective measure, but this enables direct comparisons of the values of the induced voltages at the piezoelectric electrodes in the given variation bands $\omega$. The results of the model tests are presented in the graphs (Fig. 3); they indicate that in the range of low values of the dimensionless excitation frequency $\omega \in [0,$ 6], the higher voltage values are registered in a system with a switched potential barrier. This situation occurs in relation to the relatively large amplitudes $p$ = 0.5 for the external load. If the mechanical vibrations of the object from which the energy is recovered are small, i.e. $p$ = 0.1, then the effective values of the voltage induced at the piezoelectric electrodes of the system with switched potential can be ten times higher. In the case of the impact on the energy harvesting systems with high frequencies $\omega$ > 10, the low voltage values where $u_{RMS}$ < 1 are recorded at the piezoelectric electrodes (Figs. 3a and 3c). Only in the case of the system with the three-well potential barrier is $u_{RMS}$ > 1 (Fig. 3b). The next part presents the results of the computer simulations that show the influence of the dimensionless amplitude of the mechanical vibrations $p$ on the structure of the bifurcation diagrams, which is in relation to the different locations of the minima of the two-well potential.

### *3.1. The influence of the excitation amplitude on the dynamics of the system*

From a theoretical point-of-view, the bifurcation diagrams can be plotted that use several methods. One of the most popular methods is based on the identification of the local minima and maxima of the time sequence. The same results will be obtained if the steady state diagram of a nonlinear dynamical system is identified from the phase trajectory. Nevertheless, in this case, the points that are depicted in the diagram represent the intersection of the trajectory with the displacement axis of the phase plane. In our case, the bifurcation diagrams presented in Fig. 4 are drawn based on the fixed points of the Poincaré cross-section.

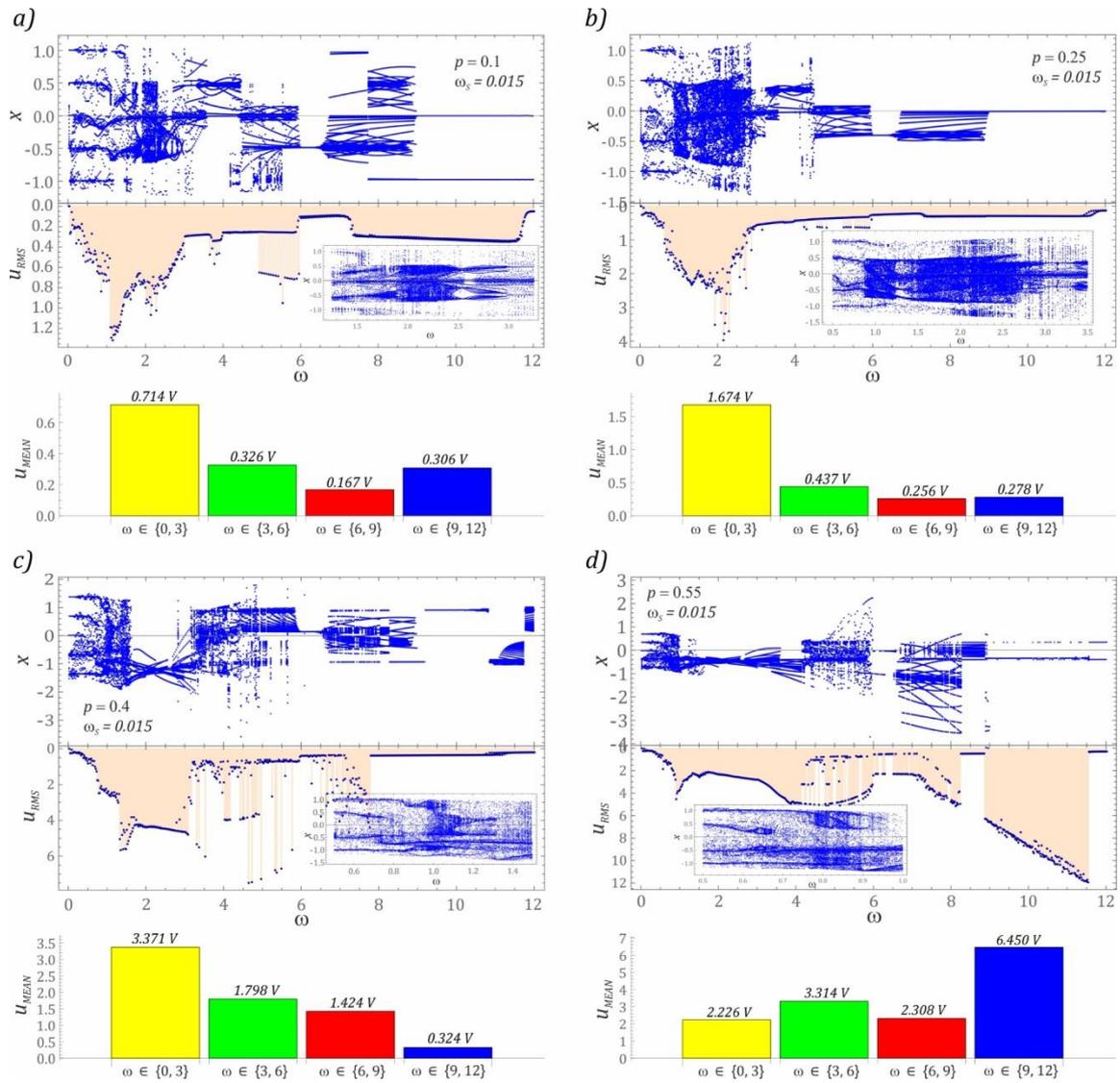

**Fig. 4.** Examples of bifurcation diagrams that show the effect of the dimensionless amplitude *p* of mechanical vibrations, which affect the energy harvesting system. Parameters of a) - d) cases are indicated in the figures

At first glance, the branches of the bifurcation diagrams are shown to resemble blurry graphic structures, which suggests the occurrence of chaotic or quasi-periodic solutions in a wide range of variability of the dimensionless frequency $\omega$. However, the periodic solutions with the relatively large periodicity are seen to dominate in the presented sample diagrams. The blurring of the diagrams, visible on magnifications, is primarily related to the presence of the transient processes that are triggered at the moments of cyclic switching of the potential barriers. Their presence can be identified by the analysis of graphical images of the time sequences for the generalized *x*-coordinate. Another reason for the appearance of the additional points on the diagram relates to the adoption of a time interval that is too short, in which the extinction of the transient processes or unstable chaotic solutions arise. It is also worth considering the presence of unstable periodic solutions [44–46], which are most often attracted to a stable periodic orbit over time. Based on the identified diagrams of the effective voltage values induced at the piezoelectric electrodes, it can be concluded that, for low values of the dimensionless amplitude of the mechanical vibrations $p < 4$, the highest voltage values are recorded in the range of the low frequencies $\omega \in [0.3]$ (Figs. 4a to 4c).

However, alongside an increase in the level of mechanical vibrations that affect the system, the harmonic components in the range of the high values $\omega < 7$ are excited (Fig. 4d). The next part presents the results of numerical experiments that provide examples of chaotic solutions; these are visualized by means of Poincaré sections, time sequences, and Fourier frequency amplitude spectra. Time events when the dynamics of the energy harvesting system are mapped with a two-well potential are highlighted with light yellow rectangles in Fig. 5. The obtained results of the computer simulations indicate that, in a system with a switched potential barrier, there is a special case of the nonlinear dynamics, which is manifested by the cyclical occurrence of chaotic and periodic solutions with a periodicity of 1T. Such behavior of the system dynamics is confirmed by the values of the estimated $D_C$ correlation dimensions and the box-counting $D_B$ dimensions of the Poincaré cross-sections, which are diametrically different. Their values suggest that the $D_C$ correlation dimension characterizes the dynamics of the energy harvesting system with switched potential barriers in the long term. From a theoretical point-of-view, we are dealing with a predictable solution because the periodic and chaotic solutions cyclically repeat.

For this reason, $D_C$ takes small values that approximately equal to zero. While $D_B$ exceeds the value of 1 relatively little, which clearly proves the presence of a chaotic solution. The system damping has a significant influence on the value of the box-counting dimension. In our simulations, we assumed a high value of the energy dissipation coefficient. This was chosen so that the transient processes caused by the switching of potential barriers would fade out relatively quickly. In the diagrams of Fig. 5a, we deal with the chaotic solutions, in which the potential is mapped with two wells. On the other hand, the examples illustrated in the graphs of Fig. 5b show the situations of a chaotic solution, in which the three-well potential barrier is active. The plotted Fourier amplitude frequency spectra clearly indicate the dominance of the harmonic component, which corresponds to the frequency of the mechanical vibrations that affect the energy harvesting system. Regardless of the nature of the solution, the identified Fourier spectra show the excitation of the harmonic components in a wide range of variability, in the band that is located below the dominant frequency. In the examples of Fig. 5b, in the vicinity of the frequency of the excitation source, the harmonics are affected that are distant from the fundamental frequency by a value equal to the frequency with which the potential barriers are switched.

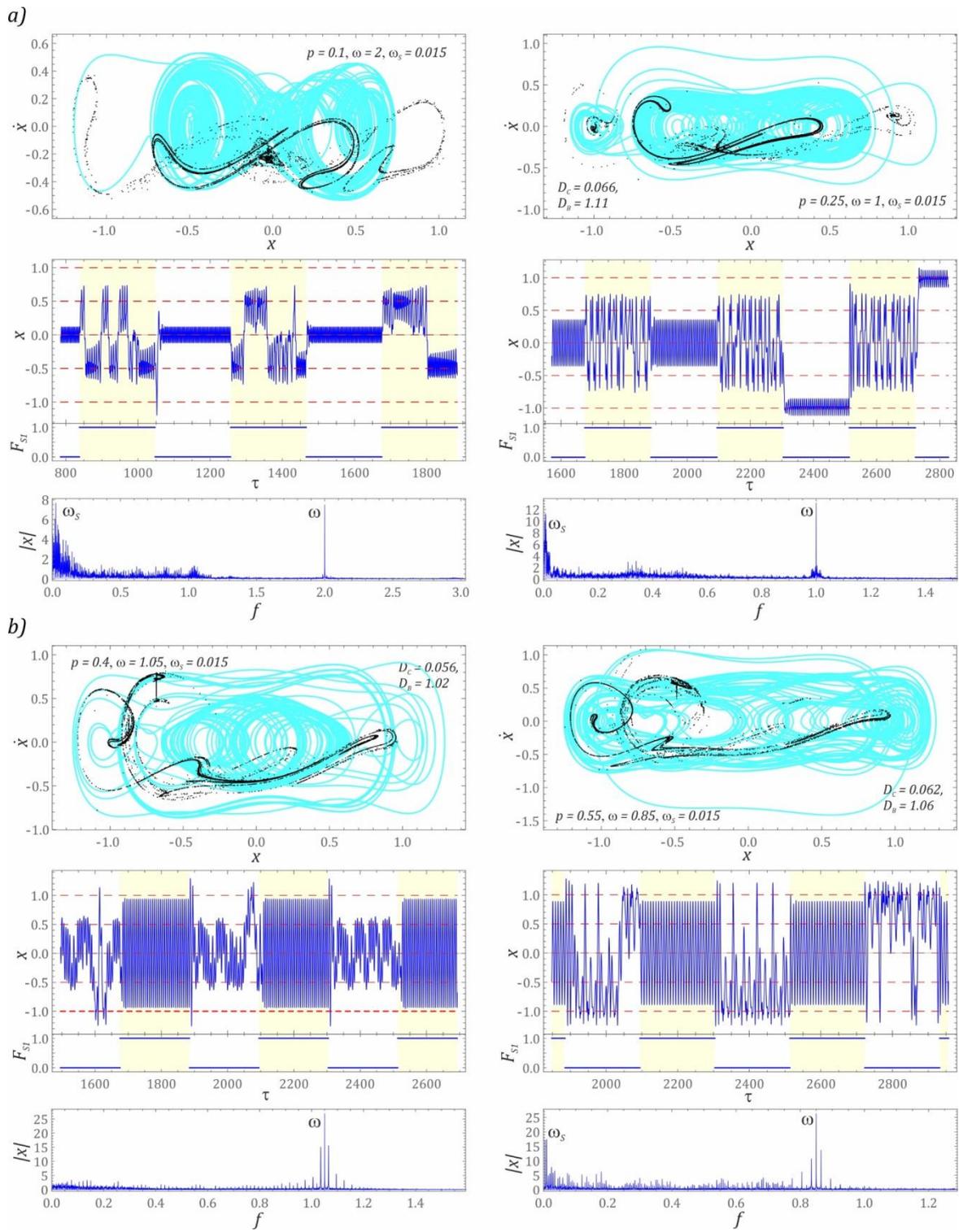

**Fig. 5.** Sample images of the chaotic solutions in terms of the corresponding time series, phase portraits, and Fourier spectra. The white and yellow backgrounds of the time series indicate the applications of alternative potentials. Parameters of a) and b) cases are indicated in the figures

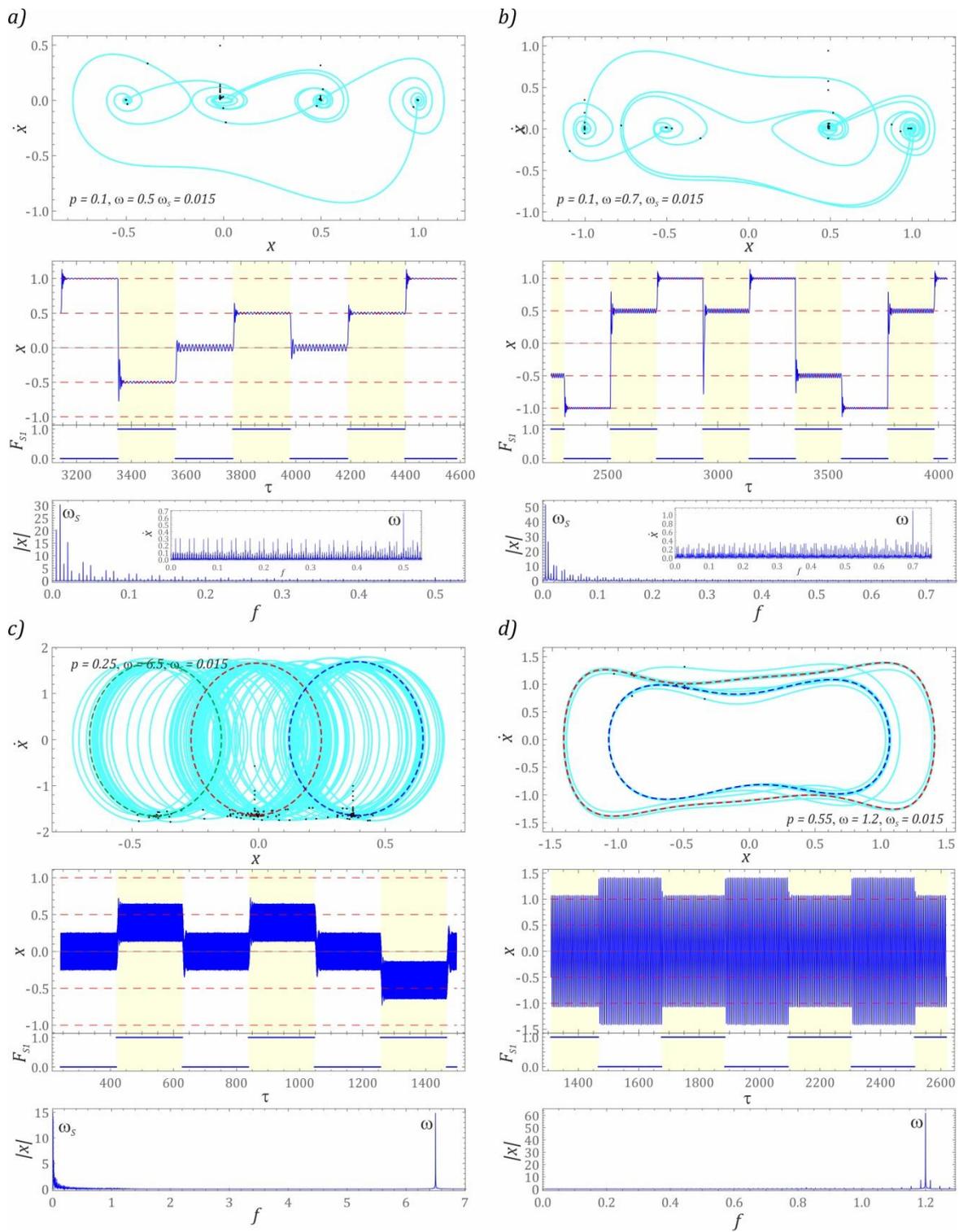

**Fig. 6.** Sample images of the periodic solutions in terms of the corresponding time series, phase portraits, and Fourier spectra. Parameters of a) – d) cases are indicated in the figures. The white and yellow backgrounds of the time series indicate the applications of the alternative potentials.

Exemplary results that illustrate the periodic responses are presented in the graphs of Fig. 6. Bearing in mind that, in the tested energy harvesting system we deal with cyclically switched potentials, the obtained results were compared to the steady states of the time windows that correspond to the interaction of the active barriers. These solutions are depicted in the form of orbits, which are drawn with dashed lines. Poincaré points are highlighted for each case on the phase trajectories. Based on the number of these points, the periodicity of the solution is identified. Regarding systems with switched potential barriers, this tool is shown to not live up to expectations. This situation occurs because there are additional points that are directly related to the transient processes, which are initiated by the switching of potentials. Therefore, all the points located outside of the curve represent the phase trajectory that should be treated as numerical artifacts.

In the range of low values of the dimensionless frequency for the mechanical vibrations $\omega$, regardless of the value of the amplitude $p$, we deal with the responses whose trajectories in the steady state are given as orbits (Fig. 6a). The relatively small changes in $\omega$ essentially affect the order (Fig. 6b) and the number of "visited" wells by the phase trajectory. In general, these denote solutions that are characterized by a relatively low energy harvesting effectiveness, which is confirmed by the diagram of the effective values of the voltage that is induced on the piezoelectric electrodes. This is the case because, despite the large orbit in the global terms, the solutions corresponding to the individual switching cycles are located inside the well. A similar behavior for the energy harvesting system also occurs in the case shown in the graphs of Fig. 6c.

Low harmonics dominate in the amplitude-frequency spectra of such solutions, which are a multiple of the frequency with which the potentials are switched (Fig. 6a). However, in relation to the higher values of the excitation frequency $\omega$ (Fig. 6c), the Fourier spectrum is dominated by the harmonic components that correspond to the frequency of excitation $\omega$ and switching potentials $\omega_s$. Alongside the level of the mechanical vibrations that affect the energy harvesting system, the zone of the solutions of this nature becomes narrower and shifted towards low values of $\omega$ (Fig. 4). With regard to the solutions characterized by the large orbits surrounding the wells of the potential barriers (Fig. 6d), in the amplitude-frequency spectra, the dominant harmonic component corresponds to the frequency of the mechanical vibrations that affect the energy harvesting system. In its vicinity, the components are excited and the values of which are reduced or increased based on the value of the potential switching frequency $\omega_S$.

### 3.2. Influence of the potentials switching frequency on the dynamics of the system

The results of the numerical experiments that illustrate the influence of the frequency $\omega_S$ of the switching potential barriers on the nature of the solution are presented below. Model tests were performed with the assumption that a constant dimensionless amplitude $p$ represents the mechanical vibrations. The results were visualized in the form of correlated bifurcation diagrams and the $u_{RMS}$ voltage induced at the piezoelectric electrodes. In addition, the effectiveness of harvesting energy for individual bands with a variability of the dimensionless frequency of the external load $\omega$ was estimated.

In the range of low switching values for the potential barriers $\omega_S < 0.1$, periodic solutions dominate in the bifurcation diagrams. Responses with a periodicity of 1T are observed in the range for the very small values of $\omega_S$. As the frequency of the potential switching increases, the single periodic solutions (Fig. 7a) evolve responses with a large or very large periodicity (Fig. 7b). It is noteworthy that the area of high frequencies for the external mechanical vibrations affect the system; in fact, we are dealing with an energy harvesting effectiveness that is practically zero, because the estimated values of the voltage $u_{RMS}$ is less than 0.2.

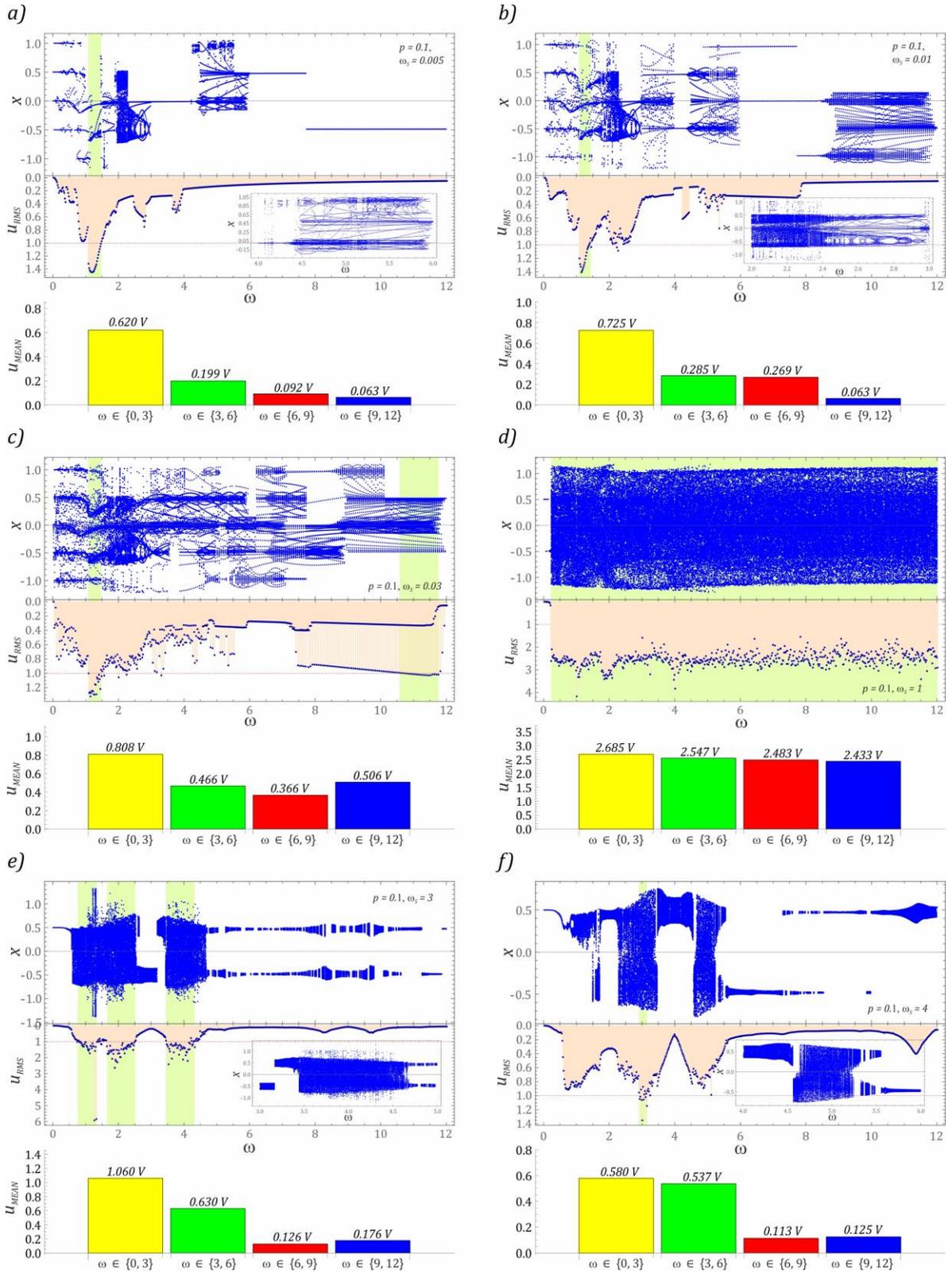

**Fig. 7.** Examples illustrating the influence of $\omega_S$ on the structure of the bifurcation diagrams for constant value of dimensionless amplitude *p=0.1*. Parameters of the corresponding cases a)-f) are indicated in the figures.

If the frequency of potential switching is equal to $\omega_S$ = 0.03, there is a much higher effectiveness for the energy harvesting in the band $\omega$ > 8 (Fig. 7c). Then, on the diagram of the $u_{RMS}$ voltage values, an extra branch appears, which indicates an additional coexisting solution, which is characterized by a high periodicity. Alongside an increasing frequency for the switching potentials $\omega_S$, the initially dominant periodic solutions with 1T periodicity evolve in response to the large or very large periodicity. This behavior for the energy harvesting system is observed until $\omega_S \approx 0.15$. From then on, the bifurcation diagram is dominated by the solutions with an unpredictable nature that arise in a wide range of variability for the dimensionless excitation frequency $\omega$ (Fig. 7d). In this case, the basically similar average voltage values are observed in the individual variability bands $\omega$. This behavior for the energy harvesting system occurs until the value of $\omega_S \approx 2.5$ is reached. A further increase in $\omega_S$ causes a decrease in the chaotic solutions and the appearance of the periodic solutions, the responses of which are limited by a two-well potential barrier (Fig. 7e). The mean values that are recorded in the individual bands of $\omega$ variation are more than two times smaller than the case depicted in the graphs of Fig. 7d.

However, a similar behavior is observed when $\omega_S$ = 4; although, the effective values of the voltage induced at the piezoelectric electrodes are much lower in this situation. For each of the cases, the results of which are presented in the graphs of the numerical experiments (Fig. 7). Here, the energy harvesting effectiveness index was determined, which was calculated as the mean $u_{MEAN}$ in a given variation band $\omega$. Moreover, the light green color distinguishes the zones in which the RMS voltage values $u_{RMS}$ are greater than 1. On this basis it was found that, regardless of the $\omega_S$ value, the highest values of the voltage induced on the piezoelectric electrodes are located in the band $\omega \in [0, 3]$; the case of $\omega_S$ = 1 is excluded from this statement. Next, exemplary images of the phase trajectories for the periodic solutions are presented (Fig. 8). If the frequencies $\omega_S$ of the potential switching are low and the dimensionless frequency of the mechanical vibrations that affect the energy harvesting system is located within the band $\omega \in [0, 1]$, then the orbits of the periodic solutions with their geometric shape are similar to the responses presented in the diagram of Fig. 6a. As for the remaining cases, solutions with lower vibration amplitudes for an elastic cantilever beam are involved. At the moment of activation of the two-well barrier, the phase trajectory is precipitated. However, with the passage of time, the transitional processes are extinguished relatively quickly. The orbits of the solutions in the temporary steady states are visualized using a dashed curve.

The expression "temporary steady states" should be understood as meaning such a dynamic state of the system, which occurs in relation to the active potential barrier. For example, in the case depicted in the graphs of Fig. 8a, the periodic orbits with a periodicity of 1T, which correspond to the temporary steady states, are located in the extreme wells of both potential barriers. Two components dominate in the amplitude-frequency spectrum, which represent the frequencies of the mechanical vibrations $\omega$ and the switching frequencies of the potentials $\omega_S$. In addition, a number of the harmonic components in their vicinity are excited, but their amplitudes are much smaller. A similar nature of the response occurs that accounts for the case shown in the graphs of Fig. 8b. Nevertheless, this time temporary solutions, also with 1T periodicity, are located in the central well and within the extreme potential barriers. The shape of the Fourier spectrum is similar to the spectrum of Fig. 8a; however, the amplitudes are much higher. Alongside the increase in the $\omega_S$ value, a limitation of the vibrations of the elastic cantilever beam is observed. Nonetheless, the orbits of the solution can assume to have complex geometric structures as well as classic oval shapes. In the case of Fig. 8c, a global periodic solution with a periodicity of 2T arises. Regarding the higher values of $\omega_S$, quasi-periodic solutions are most often used (Fig. 8d).

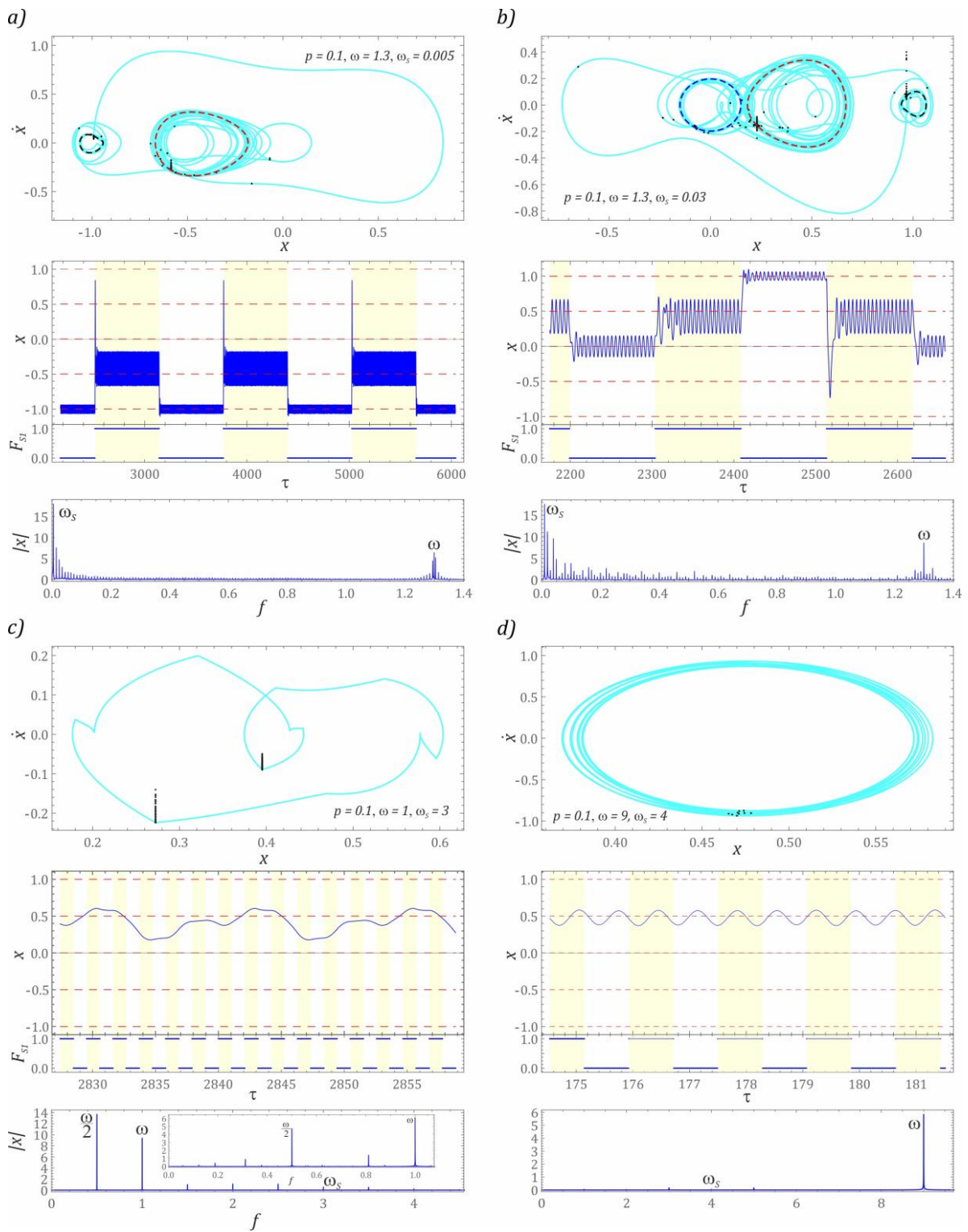

**Fig. 8.** Examples illustrating the influence of $\omega_S$ frequency on the periodic solutions in terms of the corresponding time series, phase portraits, and Fourier spectra. Parameters of a)-d) cases are indicated in the figures. The white and yellow backgrounds of the time series indicate the applications of the alternative potentials

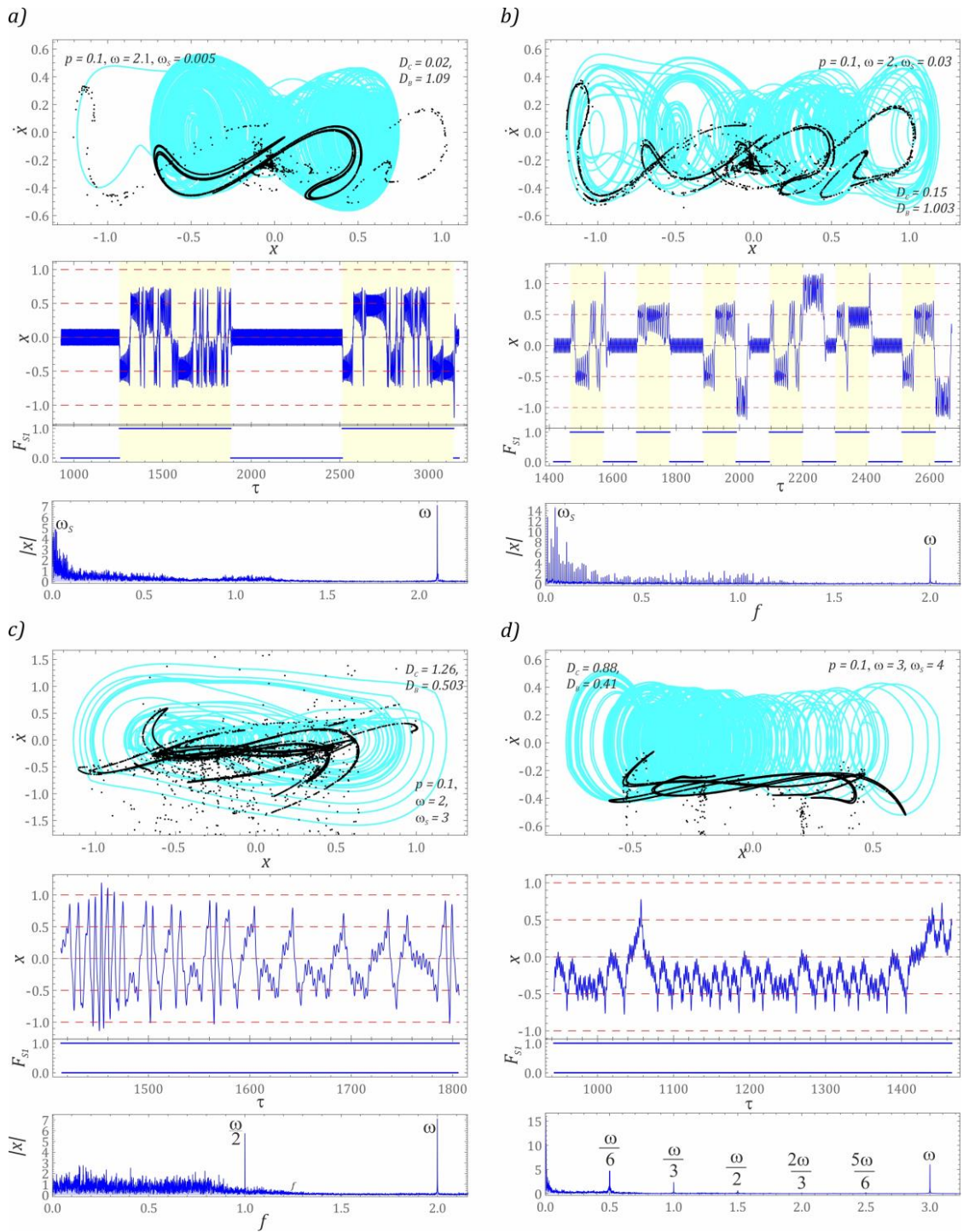

**Fig. 9.** Examples illustrating the influence of $\omega_S$ frequency on the chaotic solutions in terms of the corresponding time series, phase portraits, and Fourier spectra. Parameters of a)-d) cases are indicated in the figures. The white and yellow backgrounds of the time series indicate the applications of the alternative potentials

If the frequency of the potential switching occurs at low values $\omega_S < 0.1$, then the zones of the chaotic solutions do not change their position in principle (Figs. 7a - 7c). In the Poincaré cross-sections that are drawn, it is possible to distinguish the geometric structures multiplied and deformed by the vector field (Figs. 9a and 9b). It is also worth noting that the chaotic solutions occur when the energy harvesting system is mapped with a two-well potential. In the amplitude-frequency spectra, it is difficult to distinguish between the frequency of the excitation $\omega$ and the switching potentials $\omega_s$ for any cause-effect relationships. The behavior is different if the switching of the potential occurs quickly (Figs. 9c and 9d). Then, in the amplitude-frequency spectra, the additional harmonics are excited, the components of which are proportional to the frequency of the load that affects the energy harvesting system.

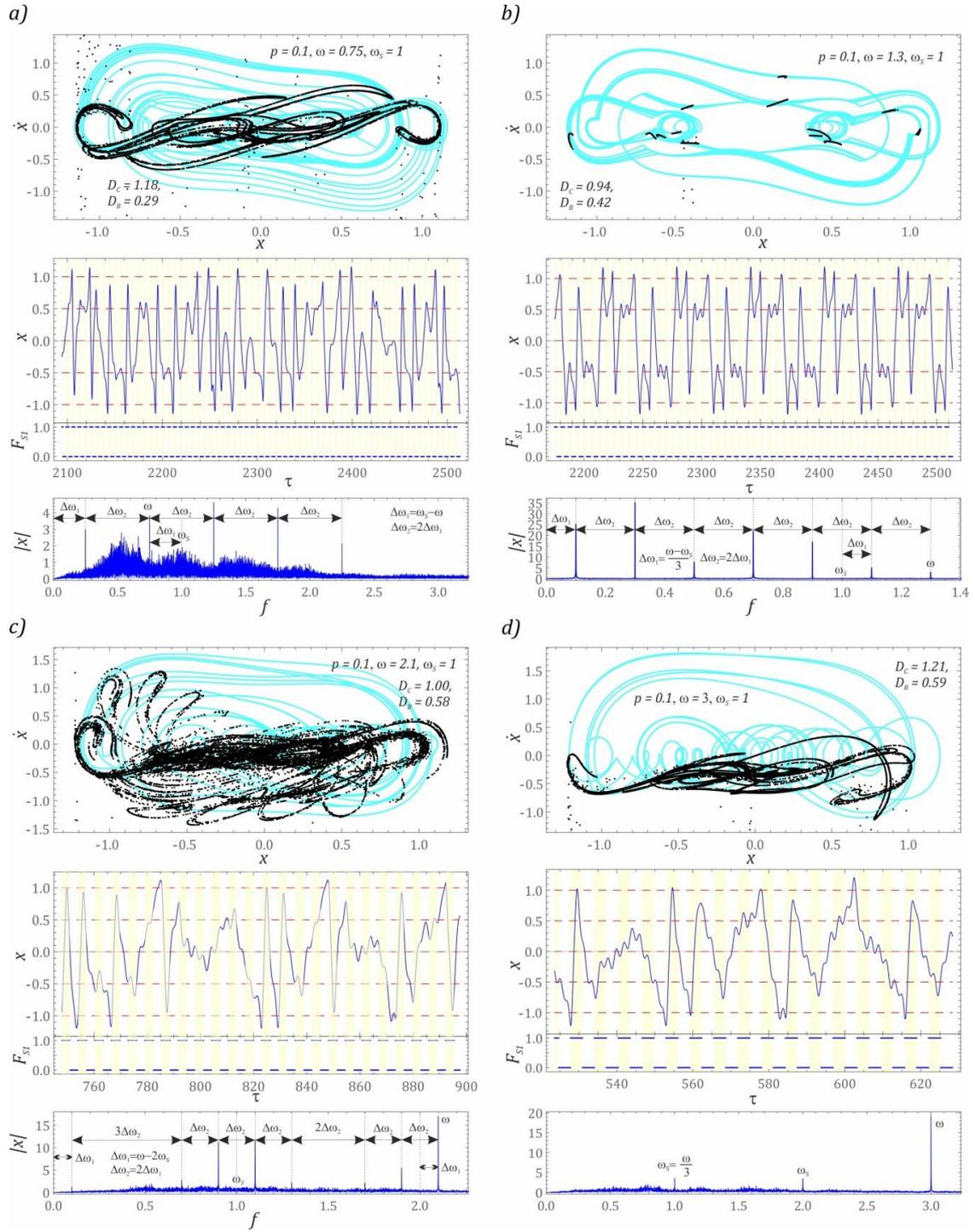

**Fig. 10.** Examples illustrating the influence of the frequency of the potential switching $\omega_S$. Parameters of a)-d) cases are indicated in the figures. The white and yellow backgrounds of the time series indicate the applications of the alternative potentials

Here we indicate that the periods of the two-well potential, which have an impact on the system, were not distinguished in the light yellow representing the time responses of the vibrations of the flexible cantilever beam. This was accomplished because the frequency of switching the barriers is so high that the $F_{S1}$ time diagram takes the form of a horizontal line. Chaotic solutions occur in a wide range of $\omega$ variability (Fig. 10), which indicate the presence of cause-and-effect relations between the frequency of the source of excitation $\omega$ and the switching of potential barriers $\omega_S$. Simultaneously, the appropriate dependencies that enable an estimation of the induced locally dominant harmonics are included in the images of the plotted Fourier spectra. It is noteworthy here that only in the case of high frequencies $\omega_S$ is the dominate components those that represent the load that affects the energy harvesting system.

### *3.3. Influence of the sequence of potential barriers switching*

One of the basic properties of the nonlinear dynamical systems is its sensitivity to changing initial conditions. This feature is manifested by the possibility of the coexistence of many solutions with regard to the same load conditions affecting a nonlinear system. With this in mind, numerical experiments were performed to assess the impact of the sequence of switching on the potential barriers. The results of the model tests presented below were related to the influence of the sequence of switching on the potentials on the geometric structure of the bifurcation diagrams (Fig. 11). By the expression "sequence of switching on", we can understand the potential characterizing the energy harvesting system at the initial moment $\tau = 0$. The results of the computer simulations presented so far have been obtained when, at the initial moment, the dynamics of the energy harvesting system is limited by the two-well $F_{S1}$ potential barrier. This case is illustrated by the potential switching cyclogram plotted in the graph of Fig. 1a. Bearing in mind which potentials are active for zero initial conditions, the $F_{Si}$ symbols are presented in the graphs.

Number 1 represents a two-well potential, while the digit 2 corresponds to the three-well potential.

In the first stage of the model research, the impact of the sequence of switching on the potentials on the structure of bifurcation diagrams, and the effective values of the voltage induced on the piezoelectric electrodes, were assessed. The obtained results of the numerical experiments indicate differences in the images of the bifurcation diagrams. However, they are noticeable in the range of higher values for the variability of the control parameter $\omega > 6$. In the range of low values of $\omega < 6$, we can basically discuss the topological similarity of the plotted diagrams. A direct comparison of the effective values of the voltage induced on the piezoelectric electrodes, and the energy harvesting in the form of average values in the individual bands of variability for the control parameter $\omega$, do not show such differences anymore. In particular, the values identified with respect to the higher potential switching frequencies (Figs. 11b and 11c) assume similar levels. It is even possible to state that they are practically the same and that the differences are within the error limit of the numerical calculations.

With regard to the low switching frequencies $\omega_s < 0.01$ (Fig. 11a), it is necessary to perform long-term numerical calculations that consider a very large number of load periods acting on the system. The operation time of computer simulations that is too short can provide the results characterizing the vibrations of a flexible cantilever beam that is realized in relation to a single potential. Similar values for the average voltages induced at the piezoelectric electrodes, in the individual bands of the variability of the control parameter $\omega$, suggest the presence of the same solutions. This is regardless of the order in which the potential barrier was activated at the initial moment. Bearing in mind the verification of the hypothesis that is formulated in this way, the responses of the periodic and chaotic solutions are presented in the following section, with reference to the same load characteristics and switching frequency.

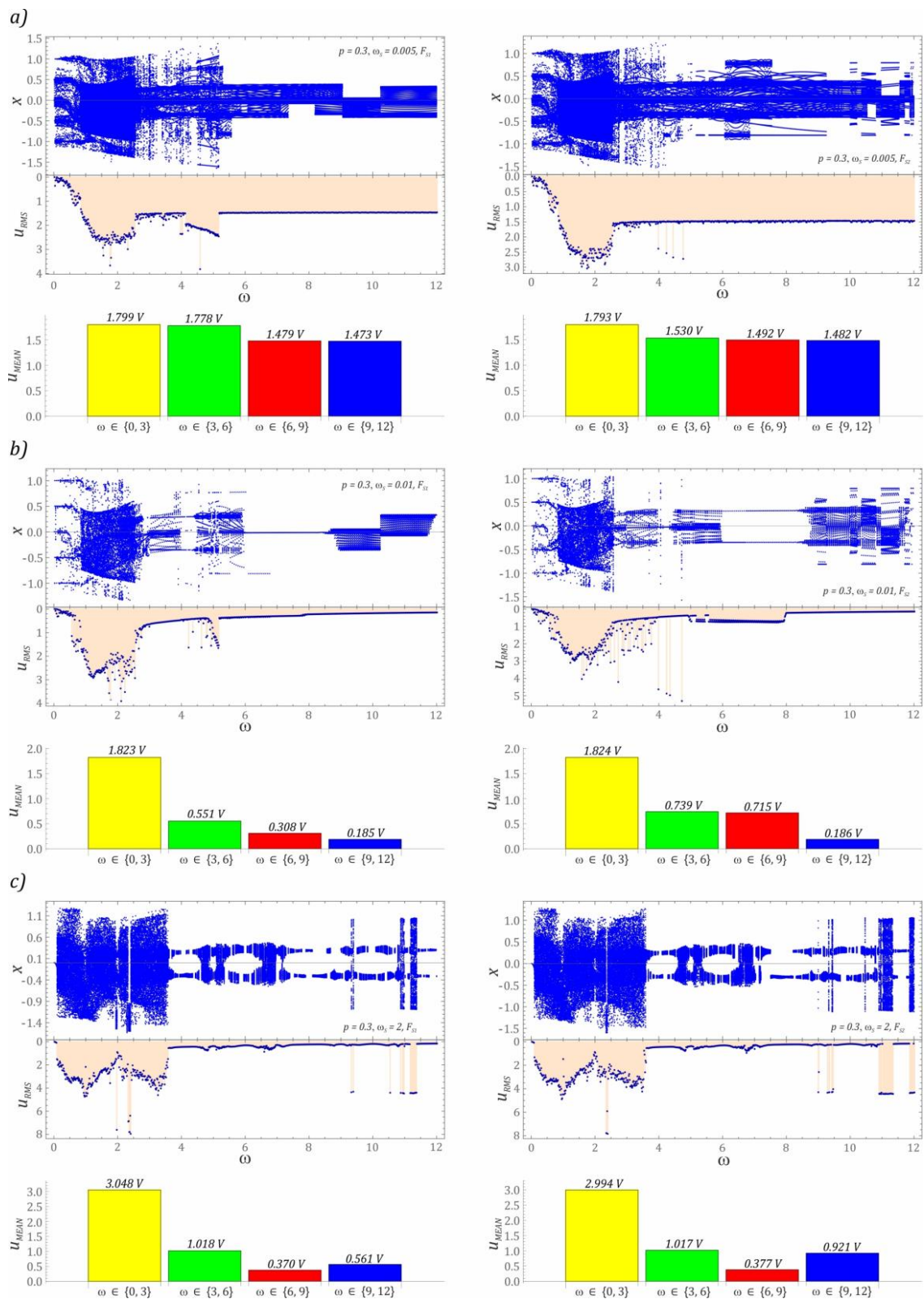

**Fig. 11.** Examples illustrating the influence of $\omega_S$ on the structure of the bifurcation diagrams for constant value of dimensionless amplitude $p=0.3$. Parameters of the corresponding cases a)-c) are indicated in the figures.

In relation to the images of the phase trajectories for the periodic solutions, the influence of the sequence of switching on the potentials is not noticeable in the range of low values of $\omega_s$ (Fig. 12a). The situation is similar with regards to the amplitude-frequency spectrum. It is noteworthy here that the character of the solution is decisively influenced by the dominant harmonic component of the Fourier spectrum, which is equal to the frequency $\omega_s$. Significant differences appear in the generalized coordinate time responses. The plotted time sequences show topological similarity, but a time shift is observed for the signal, the value of which is equal to the potential switching period. When the potentials change with a high frequency, there is virtually no time shift in the signal (Fig. 12b). The amplitude-frequency spectra also show no significant differences. On the other hand, the orbit of the solution is located in another well of the potential barrier.

In the case of the chaotic responses, the images of the phase trajectories as well as the amplitude-frequency spectra do not show significant differences. Such behavior for the solutions is observed in the case of small values of $\omega_s$ (Fig. 13a). Significant differences appear when comparing the time sequences directly. It is noteworthy that, at its basis, it is difficult to determine the size of the shift in the time sequence, as was the case for the periodic solution (Fig. 12a). These nuisances are caused by the fact that the chaotic responses de facto represent the composition of many unstable periodic solutions. For this reason, an unambiguous and precise determination of the shift of time sequences may even turn out to be impossible. A detailed analysis of the Fourier spectra shows the differences of the excited harmonic components located in the vicinity of the excitation frequency $\omega$ and the switching of the potential barrier $\omega_s$ (Fig. 13a).

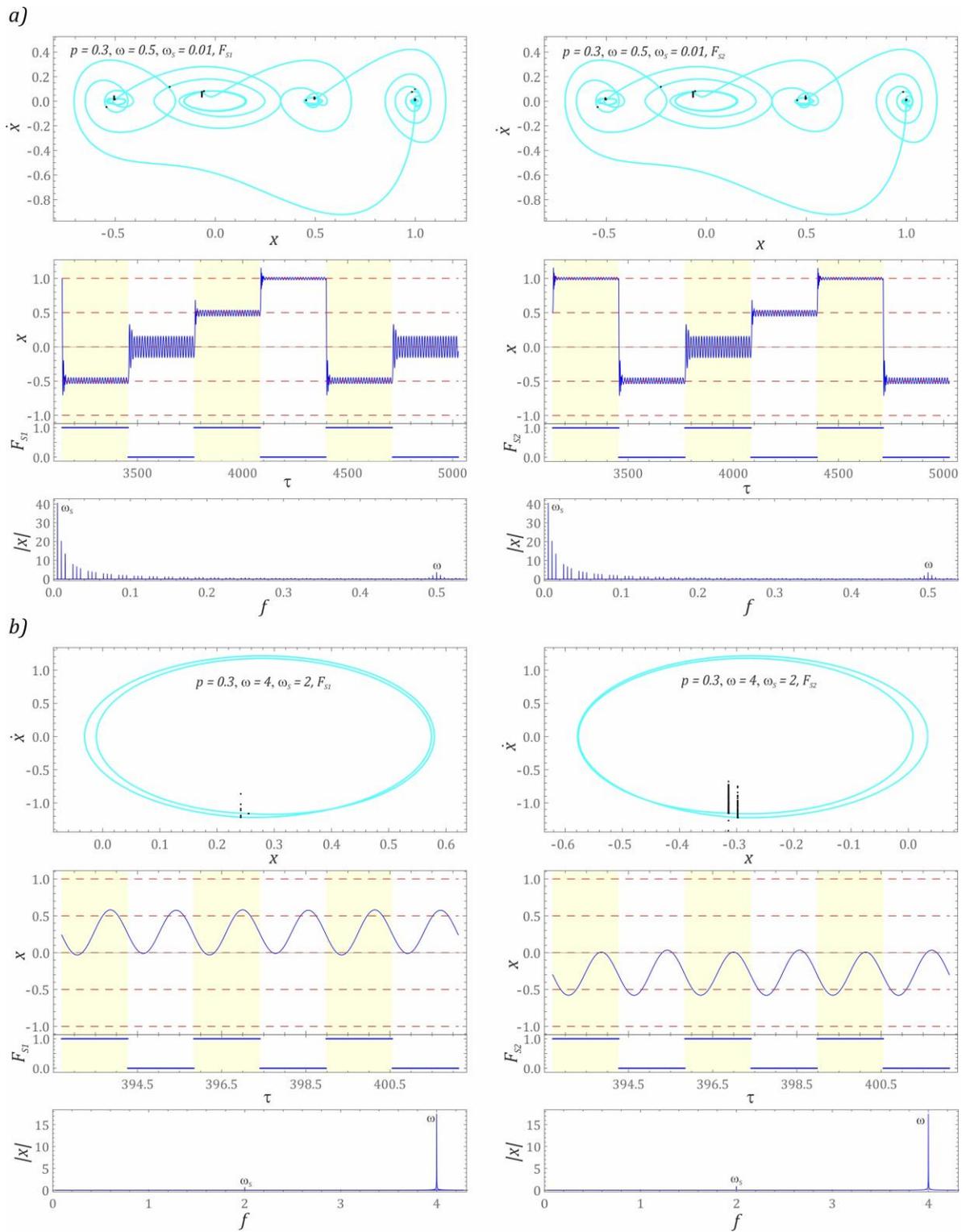

**Fig. 12.** Examples illustrating the influence of $\omega_s$ frequency on the periodic solutions in terms of the corresponding time series, phase portraits, and Fourier spectra. Parameters of a) and b) cases are indicated in the figures. The white and yellow backgrounds of the time series indicate the applications of the alternative potentials

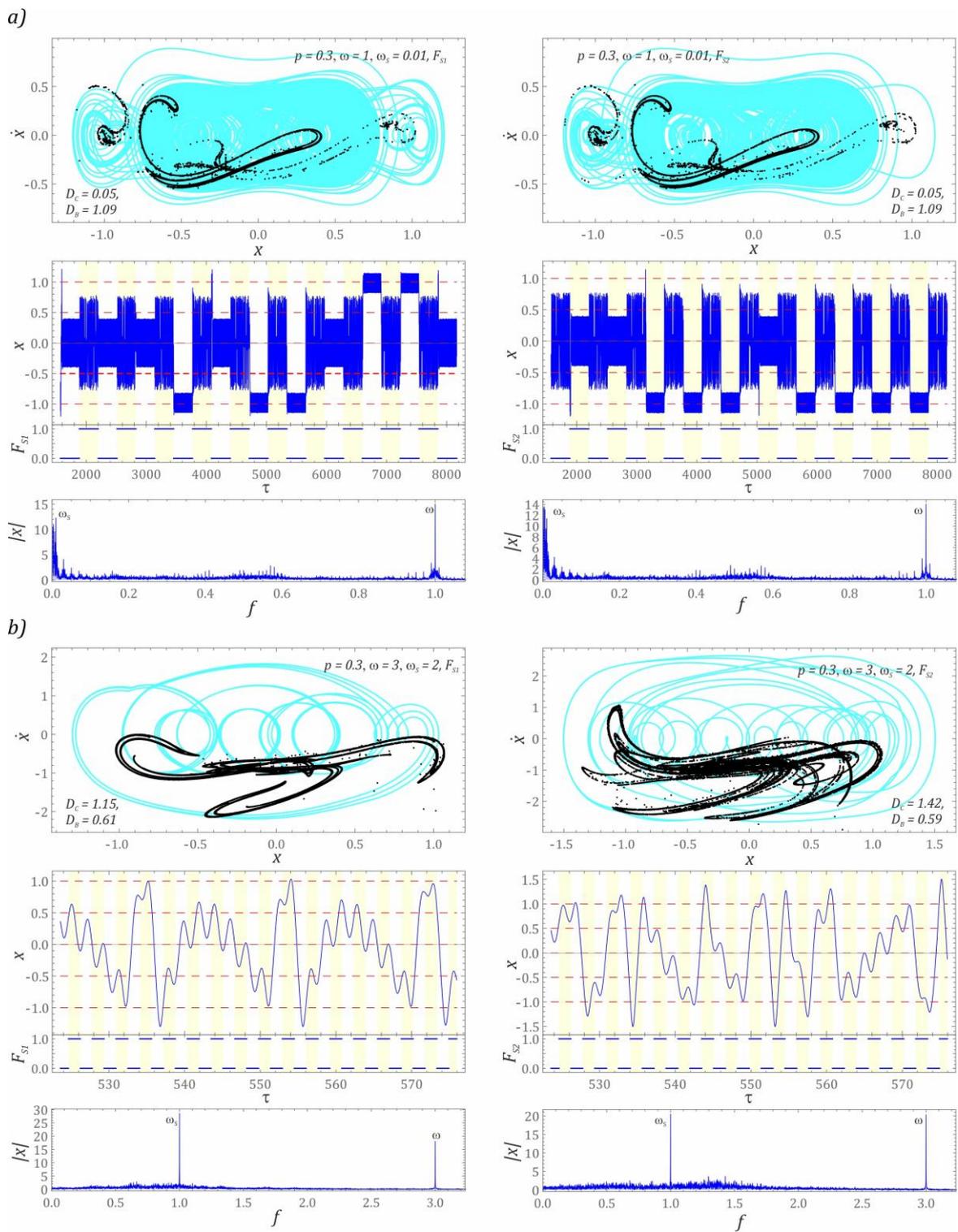

**Fig. 13.** Examples illustrating the influence of $\omega_S$ frequency on the chaotic solutions in terms of the corresponding time series, phase portraits, and Fourier spectra. Parameters of a) and b) cases are indicated in the figures. The white and yellow backgrounds of the time series indicate the applications of the alternative potentials

Significant differences occur when the potential barriers switch at high frequencies. This is especially noticeable in the images of the Poincaré cross-sections as well as the values of the correlation and box-counting dimensions (Fig. 13b). The situation is similar in relation to the drawn time series of the generalized coordinates and to the amplitude-frequency spectra. It is noteworthy that the common denominator of the harmonic Fourier spectrum distribution is the dominance of two frequencies. The values of these frequencies correspond to the mechanical vibrations that affect the energy harvesting system $\omega$ and the switching frequency of the potentials $\omega_s$. Nevertheless, if at the initial moment $\tau = 0$ the barrier is mapped with the three-well potential, the amplitudes of both harmonics are comparable. It should be clearly indicated here that such a comparison cannot be precise because the amplitudes of the Fourier spectrum are calculated with a certain accuracy. However, the two spectra differ in the bandwidth of the excited components (Fig. 13b).

## 4. Summary and final conclusions

This paper presents a detailed study on the dynamics of a new design solution for the energy harvesting system with switchable potential barriers. The presented results of the numerical experiments were limited to zero initial conditions. This approach to the published results of the computer simulations is supported by the fact that at the time $\tau = 0$ the flexible cantilever beam is in a static equilibrium position, which in fact corresponds to zero initial conditions. As a measure of system effectiveness, we adopted $U_{RMS}$ because for different resistances the measured power will also differ. It is therefore possible to easily convert the theoretical results obtained by us to a wide range of practical applications depending on the specific energy receiver. Our work is therefore a general approach possible to implement in the analysis of various types of energy harvesters. Based on the numerical experiments of the energy harvesting system with a switched potential barrier, it is possible to formulate the following conclusions:

- From the construction point-of-view, energy harvesting systems with switchable potential barriers should have large values for the damping coefficients. This is because their high values minimize the influence of the transient states caused by the switching potentials.
- In the range of low and high switching frequencies of the potentials, periodic solutions dominate in the bifurcation diagrams. With regard to a large $\omega_s$, the effective values of the voltage at the piezoelectric electrodes are significantly reduced.
- The highest effective values of the voltage induced at the piezoelectric electrodes are observed when, in the entire range of variability of the control parameter $\omega$, chaotic solutions are involved. Such solutions can be move effective in aperiodic passing the potential barriers and reaching large output voltage in piezoelectric electrodes with relatively small excitation level.
- In the range of small values of $\omega_s$, the potential characteristic that determine the dynamics of the system at the initial moment $\tau = 0$ is not significant from the point-of-view of the nature of the periodic responses. In the case of large values for $\omega_s$, the characteristic of the potential determining the dynamics of the system at the initial moment $\tau = 0$ causes the solution to "jump" to another well. With regard to the chaotic solutions, the sequence of potential activation is noticeable, especially in the range of large values for $\omega_s$.

In the next step the sequence of potential switch in the energy harvesting system will be realized in experiments. The results could be studied by fast camera and/or target tracking procedure [47,48] to reveal the nature of dynamical solution.

**Declaration of Competing Interest**

The authors declare that they have no known competing financial interests or personal relationships that could have appeared to influence the work reported in this paper.

**Acknowledgement**

This research was funded by National Science Centre, Poland under the project SHENG-2, No. 2021/40/Q/ST8/00362.

The authors thank prof. Shengxi Zhou from Northwestern Polytechnical University, Xi'an, China for fruitful discussions.